\documentclass[12pt]{iopart}
\usepackage{graphicx}
\begin{document}
\newcommand{\lvo}{LiV$_2$O$_4$}
\newcommand{\zvo}{ZnV$_2$O$_4$}
\newcommand{\lzvo}{Li$_{1-x}$Zn$_x$V$_2$O$_4$}
\newcommand{\tc}{$T_{c}$}
\newcommand{\msr}{$\mu$SR}

\jl{3}

\title[the muon Knight shift in c-LiV$_2$O$_4$]{Staggered
magnetism in LiV$_2$O$_4$ at low temperatures probed by the muon Knight shift}

\author{A. Koda\dag, R. Kadono\dag\l\footnote[3]{To
whom correspondence should be addressed.}, 
K. Ohishi\dag, S. R. Saha\dag, W. Higemoto\ddag, Y. Matsushita\P\
and Y. Ueda\P}

\address{\dag\ Institute of Materials Structure Science,
High Energy Accelerator Research Organization, Tsukuba, Ibaraki 305-0801, Japan}

\address{\l\ The Graduate University for Advanced Studies (Sokendai), 
Tsukuba, Ibaraki 305-0801, Japan}

\address{\ddag\ Advanced Science Research Center, 
Japan Atomic Energy Research Institute, Tokai, Naka, Ibaraki 319-1195, Japan}

\address{\P\ Institute for Solid State Physics, University of Tokyo, Kashiwa, Chiba 277-8581, Japan}

\begin{abstract}
We report on the muon Knight shift measurement in single crystals of \lvo.
Contrary to what is anticipated for the heavy-fermion state based on the Kondo 
mechanism, the presence of inhomogeneous local magnetic moments is 
demonstrated by the broad distribution of the Knight shift at temperatures
well below the presumed ``Kondo temperature" ($T^*\simeq 30$ K). 
Moreover, a significant fraction ($\simeq10$ \%) of the specimen gives 
rise to a second component which is virtually non-magnetic.  These observations
strongly suggest that the anomalous properties of \lvo\ originates from
frustration of local magnetic moments.
\end{abstract}



\vspace {1cm}

As one of the few spinel oxides which exhibit metallic conductivity, 
\lvo\ provides unique opportunity to study the potential relation
between transport property and magnetic frustration; the magnetic
vanadium ions (nominally V$^{3.5+}$, 
3$d^{1.5}$) form a three dimensional network of 
corner-shared tetrahedra (pyrochlore sublattice),
where antiferromagnetic (AF) correlation between the nearest neighboring
(nn) vanadium moments leads to a highly degenerate magnetic ground state
due to geometrical frustration.  Emergence of spin glass-like ground state
upon the substitution of Li by Zn strongly suggests that such frustration is
indeed in effect\cite{Ueda:97,Trinkl:00,Urano:00}.
In such a situation, it is predicted  that
an exotic ground state like resonating valence bond (RVB)
state might occur in the system  with small spins (e.g., $S=1/2$) where
quantum fluctuation is important\cite{Anderson:73}.  It is well known that
the RVB state has been proposed as a possible ground state for the
high-$T_{\rm c}$ cuprate superconductors\cite{Anderson:87}.
Another interesting feature is that there might be a frustration (degeneracy) of
charge state in view of nearly localized electrons; there are a variety of
mappings for the V$^{3+}$ and V$^{4+}$ ions which are
energetically equivalent on the pyrochlore lattice.

Meanwhile, the recent revelation of a heavy fermion-like behavior 
in \lvo\ has stimulated
renewed interest in this compound, where the large Sommerfeld coefficient
at low temperatures ($\gamma\simeq0.42$ J/mol K$^2$) comparable to that of 
typical $f$-electron 
heavy fermion compounds, e.g, UPt$_3$, has been attributed to the formation of
 $d$-electron 
heavy quasiparticles\cite{Urano:00,SKondo:97,SKondo:99}.
Besides the Sommerfeld coefficient, other bulk properties including spin 
susceptibility, resistivity, and the Hall coefficient commonly suggest 
a characteristic temperature, $T^*\simeq 30$ K, where the formation of such a
coherent fermionic state seems to set in.

It is established in the $f$-electron heavy fermion systems,  
particularly in those belong to the class of ``dense Kondo" system,
that the heavy quasiparticles
are formed by the Kondo coupling between conduction electrons and 
local $f$-electrons, where the local spin degree of freedom is convoluted
into the effective mass of conduction electrons through the spin-singlet
formation.  Therefore, provided
that a similar electronic state is realized in \lvo, there would be no 
possibility for the geometrical frustration to play a role in the electronic 
property; the ground state would be a uniform metallic state without 
local magnetic moments.  However, it is not obvious whether or not 
such heavy quasiparticle state is possible in transition metal oxides,
because the metallic conduction is carried by $d$-electrons which
exhibit strong local electronic correlation.

In this brief communication, we report on our recent measurement of
the muon Knight shift in single crystalline ($sc$-) \lvo.  In the previous study 
on the powder specimen, we have shown that there are two different
magnetic sectors which are largely different in their dynamical 
properties\cite{Koda:04}.  In particular, one of those exhibits a
relatively large hyperfine coupling with broad distribution at lower
temperatures which is fluctuating by a characteristic frequency of
$10^9$--$10^{12}$ s$^{-1}$.  Here we show that the fractional yield
of this magnetically fluctuating component is further enhanced in 
the single crystalline specimen, while the non-magnetic sector 
still remains as a minority component.  This observation clearly
demonstrates that the presence of local moments is an intrinsic 
property of \lvo, and thereby it strongly suggests that the excess
entropy is carried mostly by the frustrated local spins at lower 
temperatures.


It is often found in the Li compounds that the specimen exhibits rapid 
deterioration due to the migration of Li atoms and subsequent 
oxidization at the surface.  This can be avoided by the use of large
single crystal to minimize the surface area.  However, the crystal
growth of \lvo\ is quite limited due to chemical difficulties; so far it is only
the hydrothermal method that has been successful in the 
growth of $sc$-\lvo\cite{Rogers:67}.  Unfortunately, the hydrothermal
method does not fit for large and high-yield crystal growth in laboratory
because of  the difficulty to optimize conditions. Moreover, it has a 
higher possibility of Li deficiency in the crystal.  We have succeeded in 
growing single crystals with a dimension of $\sim$1 mm$^3$
using LiCl-Li$_2$MoO$_4$-LiBO$_2$ system as a solvent for 
\lvo\cite{Matsushita:05}.  The obtained crystals exhibit bulk properties
which are in perfect agreement with those reported earlier\cite{Urano:00}. 
In order to avoid the deterioration due to exposure to ambient condition in transit, 
the crystalline specimen (consisting of small crystals with net weight of $\sim$0.1 g)
was encapsulated into an evacuated vial immediately after the growth.

\msr\ measurements under a transverse field ($\simeq10$ kOe)
have been performed on the M15 beamline of TRIUMF using the HiTime spectrometer.  
The crystal specimen was loaded onto a He gas-exchange type
cryostat, and irradiated by a muon beam with a momentum of 29 MeV/c;
the muon stopping range is $\sim0.5$ mm at this momentum and thereby muons
probe the bulk crystal properties. A precaution was taken to minimize the
duration of exposure to the air while loading the specimen from the vial.  
A low-background decay-positron counter
system was employed to reduce the positron events from muons which missed
the specimen.  The relative background yield is expected to be less than 
$\sim$5 \% in the typical condition. 

\begin{figure}[t]
\begin{center}
\includegraphics[width=0.85\linewidth]{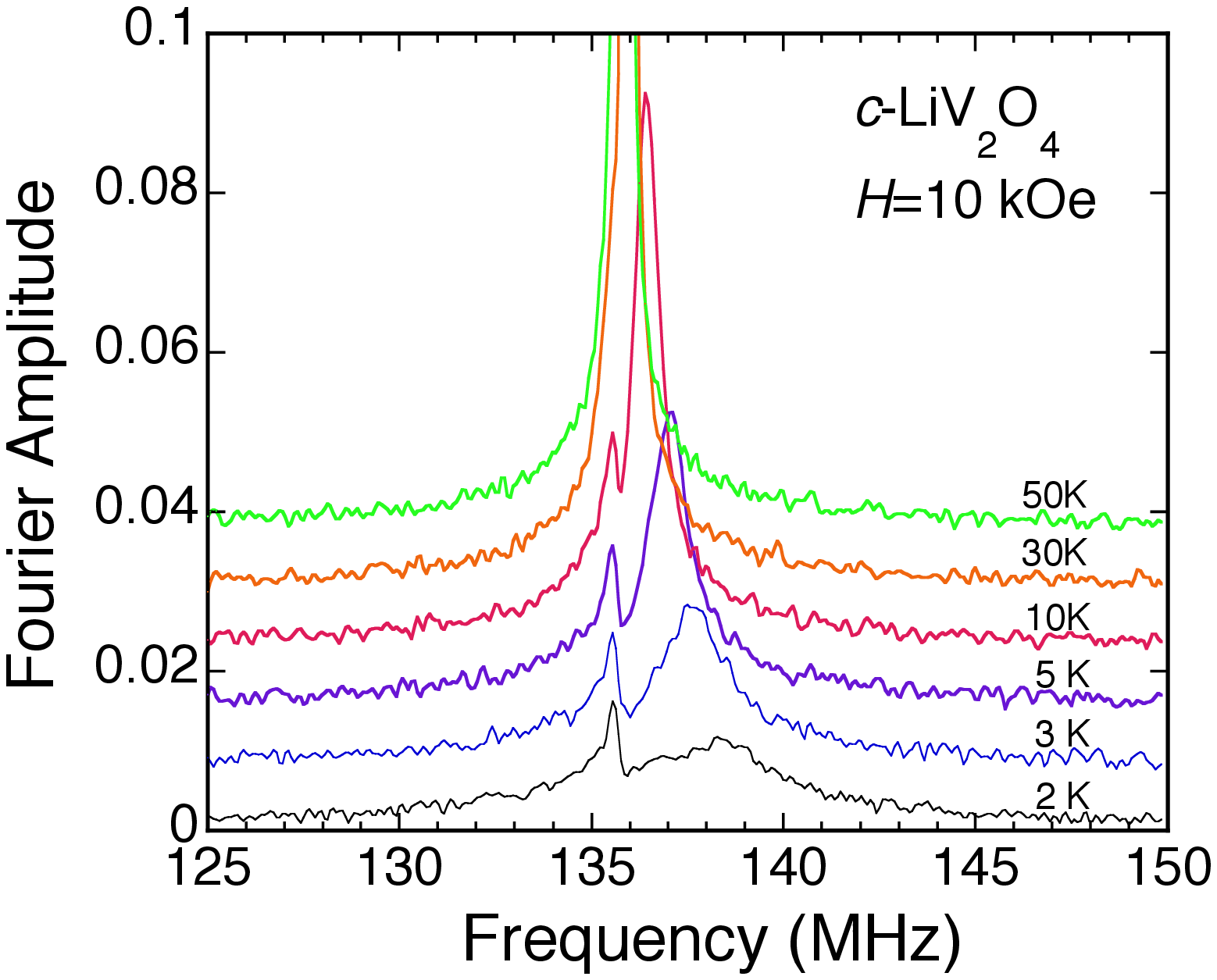}
\caption{\label{fftspec} Fast Fourier transform of the \msr\
time spectra in crystalline \lvo\ at low temperatures.
The peak near 135 MHz remains sharp, while another peak at
higher frequency becomes broad with decreasing temperature.}
\end{center}
\end{figure}

The fast Fourier transform of the \msr\ time spectra at low temperatures
is shown in Figure \ref{fftspec},
where a double peak structure is clearly observed below 10 K; detailed examination
indicates that the structure due to the satellite peak at lower frequency is visible 
already at 50 K.  While the position and 
linewidth of the satellite peak (near 135 MHz) is mostly independent
of temperature, the main peak exhibits a shift to higher frequency 
and associated increase
of the linewidth with decreasing temperature, indicating that the latter has
a much stronger hyperfine (HF) coupling with considerable distribution of the
coupling parameters.
These features are qualitatively in perfect agreement
with those observed previously in the powder specimen\cite{Koda:04}.
It is notable, however, that a considerable shift of the signal weight from
the former to the latter component is observed in $sc$-\lvo.
Moreover, the linewidth of the latter component is considerably larger
than that in the powder specimen.
The fitting analysis in the time domain using the form,
\begin{equation}
A_0G_{xy}(t)=A_0e^{-i\phi}\sum_{i=1}^n f_i\exp(i\omega_i t-\Lambda_i t),
\label{asyfit}
\end{equation}
indicates that the spectra is well reproduced by assuming three frequency
components ($n=3$), where the main peak is further split into a doublet;
here $A_0$ is the muon-positron decay asymmetry, $f_i$ is the fractional
yield of the component ($\sum f_i=1$)
with a frequency $\omega_i$ and relaxation rate $\Lambda_i$,
and $\phi$ is the initial phase.  

\begin{figure}[t]
\begin{center}
\includegraphics[width=0.85\linewidth]{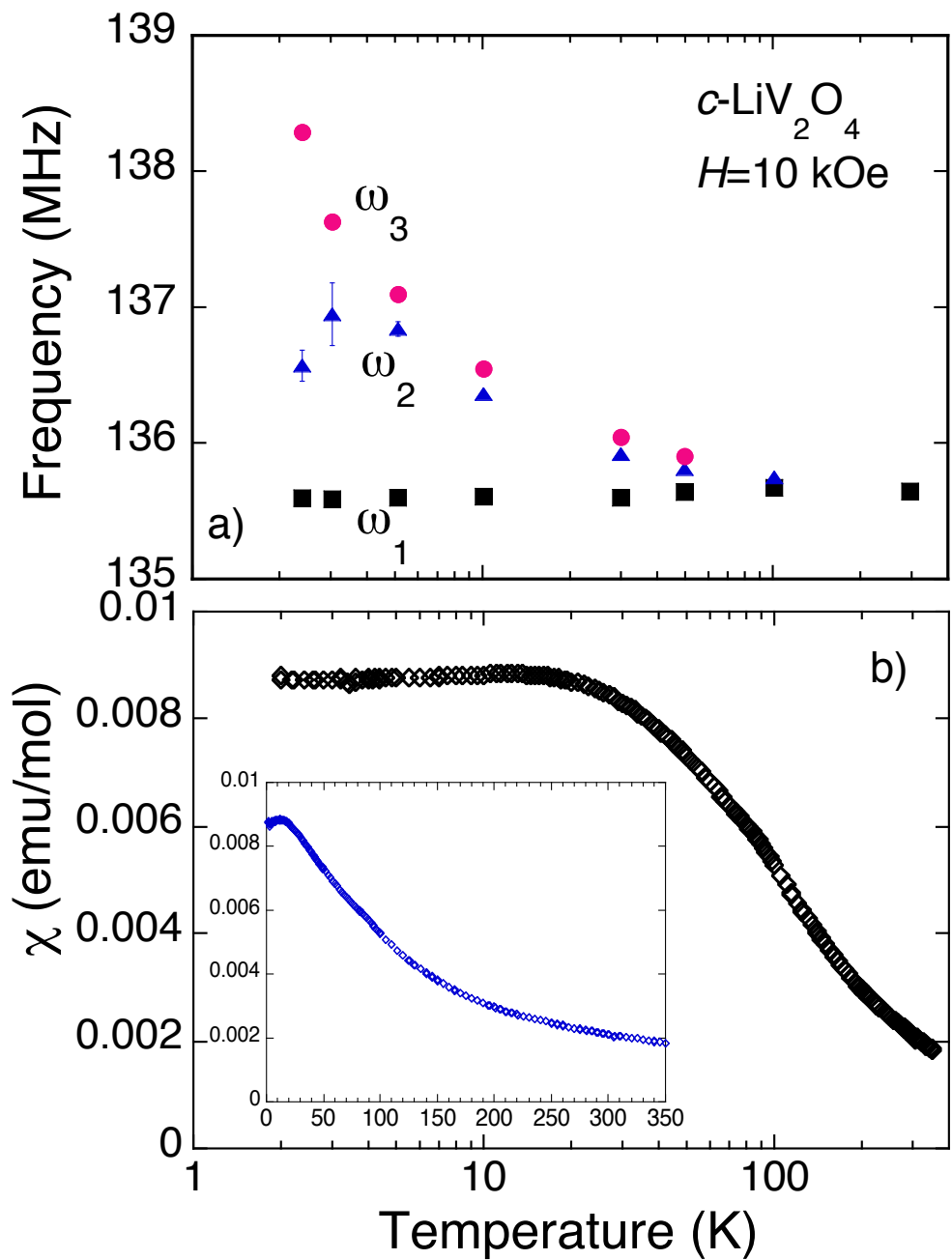}
\caption{\label{frqchi} (a) The frequency of three components vs temperature
obtained by fitting analysis
using equation (\ref{asyfit}), where the axis of abscissas is in the logarithmic scale. 
(b) The temperature dependence of magnetization in the present specimen.
Inset; the same data are displayed in a linear scale.}
\end{center}
\end{figure}

\begin{figure}[t]
\begin{center}
\includegraphics[width=0.85\linewidth]{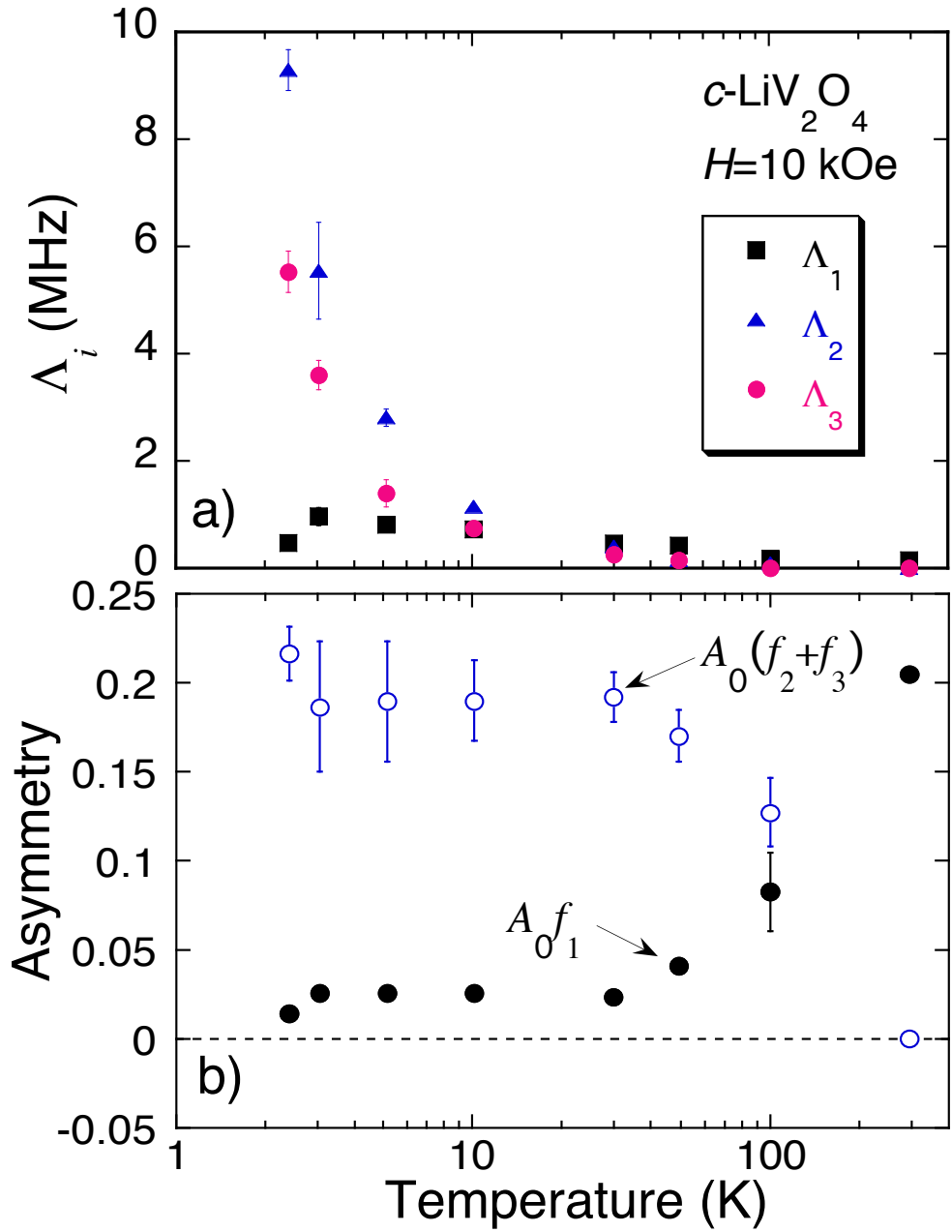}
\caption{\label{asyrlx} (a) The spin relaxation rate and (b) partial asymmetry
of three components vs temperature obtained by fitting analysis
using equation (\ref{asyfit}), where the sum $A_0(f_2+f_3)$ 
in (b) corresponds to the net asymmetry for the broad peak.}
\end{center}
\end{figure}

Figure \ref{frqchi}(a) shows the temperature dependence of the peak frequency
$\omega_i$ with the temperature plotted in the logarithmic scale.  Compared with that
of the bulk susceptibility ($\chi$) shown in Figure \ref{frqchi}(b), 
$\omega_2$ and $\omega_3$ (corresponding to the broad peak in Figure \ref{fftspec})
exhibit steep increase
over the temperature region below $T^*\sim30$ K where the susceptibility
levels off.  Here, we stress that the temperature dependence
of the bulk $\chi$ is in excellent agreement with that reported previously
for single crystal specimen\cite{Urano:00}.
The behavior of $\omega_{2,3}$ indicates that the local susceptibility probed by 
the muon Knight
shift is not proportional to the bulk susceptibility below $T^*$. 
In accordance with the muon Knight shift, the spin relaxation rate of the broad peak
increases steeply with decreasing temperature (see Figure \ref{asyrlx}(a)).
It is known from the previous \msr\ experiment under a longitudinal magnetic
field that the longitudinal spin relaxation is slow over the relevant temperature region
due to the fast fluctuation of HF fields\cite{Koda:04}.  Therefore, the broad linewidth
under a transverse field can be attributed to the distribution of HF parameters,
which is naturally expected for the case of staggered magnetism.

In contrast to the behavior of frequency shift which is markedly different above
and below $T^*$, the temperature dependence of partial asymmetry for the broad peak 
(i.e., $A_0[f_2+f_3]$ in Figure \ref{asyrlx}(b)) resembles that of $\chi$; it develops
gradually below ambient temperature and levels off around $T^*$.  
The fractional yield of the broad peak 
($f_2+f_3$) at low temperatures is about 90 \%, which is considerably 
larger than
that observed in the powder specimen ($\sim50$ \%)\cite{Koda:04}. 
Considering the quality of the present specimen, this result strongly suggests 
that the bulk electronic property is predominantly determined by the staggered
magnetism.
On the other hand, while the yield of the satellite peak ($f_1$) exhibits gradual
decrease with decreasing temperature, it stays near $\sim$10 \% at low 
temperatures.  This is considerably larger than the background level of the 
experimental apparatus, and thereby we attribute this component to the signal 
from the specimen.  It is interesting to note that such partitioning of \msr\ signal into
two components is commonly observed in Zn-doped \lvo\cite{Koda:04,Kalvius:03}.
The significant difference in the relative yield of 
those components between powder and single crystalline specimen strongly 
suggests that the magnetic
property of \lvo\ is sensitive to the quality of the specimen and that all the previous
results based on powder specimen should be re-examined in this 
respect\cite{Kaps:01,Johnston:05}.  

\begin{figure}[t]
\begin{center}
\includegraphics[width=0.85\linewidth]{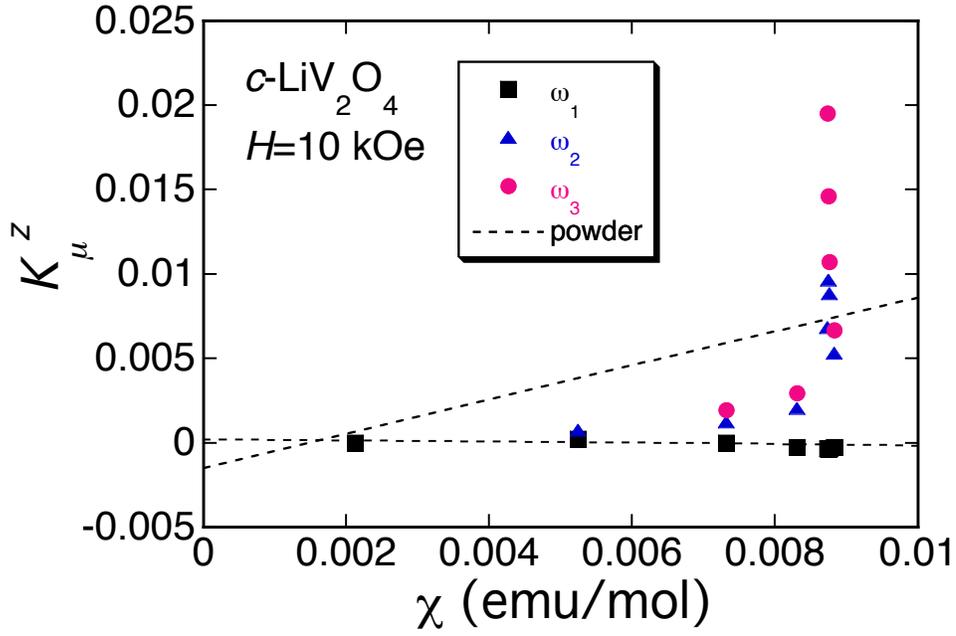}
\caption{\label{kchi} The muon Knight shift vs magnetic susceptibility
($K$-$\chi$ plot) obtained from the data in Figure \ref{frqchi}.
The dashed lines indicate the previous result on the powder specimen,
where $A_\mu^z=+5.57$ kOe/$\mu_B$ and $-0.27$ kOe/$\mu_B$
(after \cite{Koda:04}).}
\end{center}
\end{figure}

In order to examine the HF parameters in more detail, the muon Knight shift
($K_\mu^z$) is
plotted against the bulk magnetic susceptibility in Figure \ref{kchi}.
Note that the gradient in this $K$-$\chi$ plot provides 
the muon HF parameter, 
\begin{equation}
A_\mu^z=N_{\rm A}\mu_{\rm B}\frac{dK_\mu^z}{d\chi}\:,
\end{equation}
where $N_{\rm A}$ is the Avogadro number and $\mu_{\rm B}$ is the Bohr
magneton.
As is anticipated from Figure \ref{frqchi}, the $K$-$\chi$ relation is 
highly non-linear for the broad components (corresponding to $\omega_2$
and $\omega_3$), while that for the satellite line ($\omega_1$)
exhibits a linear relation
with small negative gradient corresponding to $A_\mu^z\simeq-0.40$ 
kOe/$\mu_{\rm B}$.  This is in good contrast to the typical behavior
observed in conventional heavy fermion compounds, e.g., CeRu$_2$Si$_2$
($T^*\sim10$ K)\cite{Amato:97}, where no such anomaly is observed.
In terms of the broad component, the previous result obtained 
for the powder specimen\cite{Koda:04} (dashed lines in Figure \ref{kchi})
seems to be a coarse average
of the behavior observed in $sc$-\lvo.  Thus, a broad distribution 
of HF parameter below $T^*$ is confirmed from 
this non-linear behavior of the $K$-$\chi$
plot.  In the meantime, the calculated value of $A_\mu^z$ shows reasonable
agreement with the mean value of $A_\mu^z$ ($\sim6$ kOe/$\mu_{\rm B}$) 
at the lowest temperature when the muon site is assumed to be slightly 
off the center of a ring consisting of six V$^{4+}$ ions (with their full moment
in effect) to form muon-oxygen
bond ($\simeq1$ \AA); the reason not to consider the contribution
of V$^{3+}$ is discussed below.

\begin{figure}[t]
\begin{center}
\includegraphics[width=0.85\linewidth]{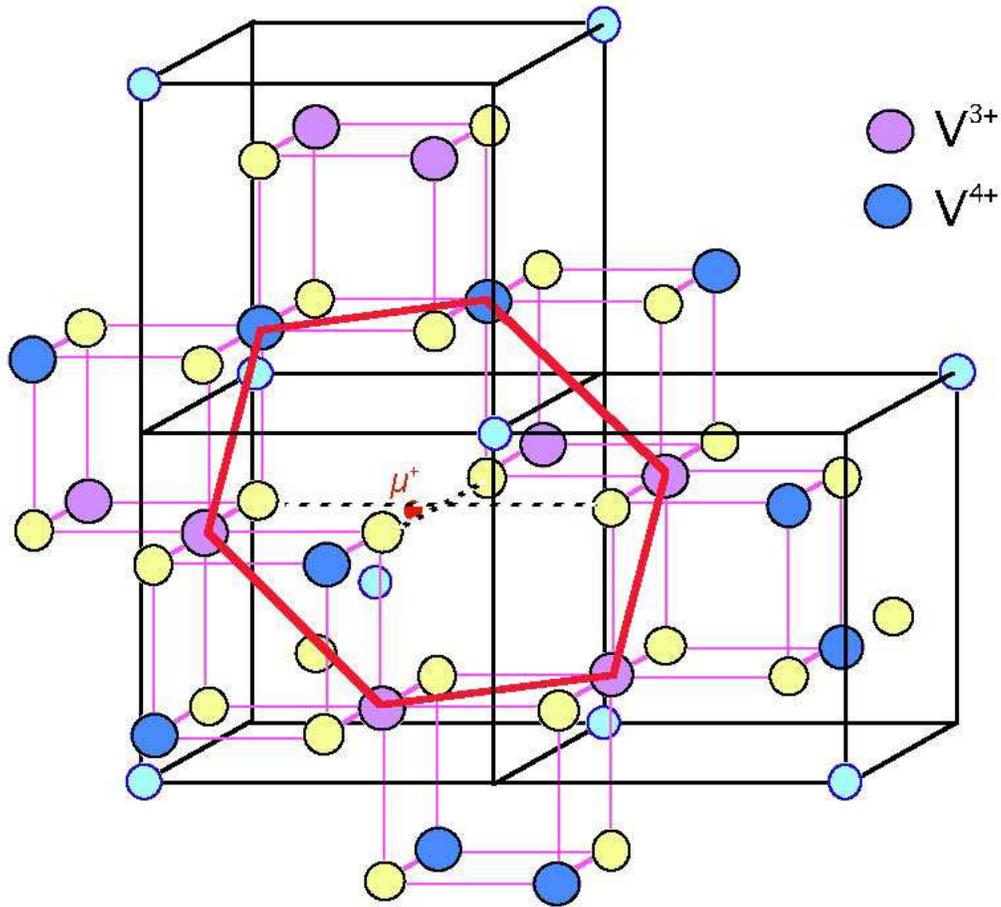}
\caption{\label{lvosite} One of the possible configurations of V$^{3+}$
and V$^{4+}$ ions which satisfy the rule that every vanadium tetrahedrons have 
two of them each. Muons are likely to be situated near the center
of octahedron cage made of oxygen atoms. The six nearest neighboring vanadium
ions are marked by a solid hexagon, where four of them belong to a Haldane chain
(consisting of V$^{3+}$ ions) and the rest to a Heisenberg chain.}
\end{center}
\end{figure}

It is inferred from the large HF parameter as well as its broad distribution
($\Delta A_\mu^z\sim 3$ kOe/$\mu_{\rm B}$) observed for the predominant
component that there are staggered magnetic moments of vanadium ions
in \lvo\ at temperatures well below the presumed Kondo temperature ($\sim T^*$).
This is quite different from what is expected for the conventional Kondo lattice
system where the local moments are quenched by the Kondo coupling to the
conduction electrons.   Thus, the present \msr\ result is
clearly inconsistent with the theoretical models which assume the
disappearance of local moments due to the Kondo effect or other 
mechanisms\cite{Anisimov:99,Singh:99,Kusunose:00,Fujimoto:02,Hopkinson:02}.
On the contrary, it strongly
suggests that the heavy fermion-like behavior observed in the bulk electronic
property is superficial, and that it is due to
the macroscopic degeneracy of the states associated
with the geometrical frustration of local vanadium moments\cite{Ogata:02}.
Along this line, a number of theoretical attempts have been made to understand
the true ground state of \lvo\cite{Eyert:99,Fulde:01,Lacroix:01,Shannon:02,Laad:03}.
Their conjecture is based on the nearly localized V$^{3+}$ 
and V$^{4+}$ ions which undergo valence fluctuation; it is interesting to note that
the situation, in terms of the charge state, 
is similar to that considered by Verwey for magnetite
(Fe$_3$O$_4$)\cite{Verwey:41}.   
Then, considering the Anderson's argument that
the two by two occupation of V$^{3+}$ and V$^{4+}$ ions on a tetrahedron 
is most favorable in terms of free energy\cite{Anderson:56}, the entire
pyrochlore lattice can be viewed as a cluster of one dimensional chains/rings
consisting only of V$^{3+}$ or V$^{4+}$\cite{Fulde:01}.  An example of
such a situation is illustrated in Figure \ref{lvosite}.
Here, those of trivalent vanadium
ions ($S=1$) correspond to the Haldane chains, and thus they would fall into
the spin-singlet state.  Meanwhile, those of tetravalent vanadium
ions ($S=1/2$) would give rise to the Heisenberg spin chains of local moments
which would remain paramagnetic at low temperatures.  It is presumable that 
the latter is responsible for the highly enhanced Sommerfeld coefficient in \lvo\
due to the entropy associated with the staggered moments\cite{Ogata:02}, 
while the former is predicted to have the Haldane gap and thereby would not contribute
to the low energy excitation.  

We point out that the observed \msr\ signals
are perfectly in line with the above mentioned ground state; the sharp satellite 
peak (which is virtually non-magnetic) might be the signal coming from the muons 
surrounded by the Haldane chains, and the broad peaks might be from those near the 
Heisenberg spin chains.  The small hyperfine parameter for the former is
attributed to the dipolar coupling between muons and paramagnetic spins at the 
edge of the Haldane chains\cite{Hagiwara:90,Kojima:95}, which would exhibit 
normal $K$-$\chi$ behavior.  On the other hand, the linewidth for the latter 
would be controlled by the valence (charge) fluctuation rate $\nu_{\rm C}$; when 
$\nu_{\rm C}$ falls
below $10^9$ s$^{-1}$, the difference in the configuration of V$^{3+}$/ V$^{4+}$ ions 
around the muon sites would lead to the distribution of hyperfine parameters
$\Delta A_\mu^z$ because 
the time-averaging mechanism becomes ineffective.  Here, we speculate 
that $\nu_{\rm C}\le 10^9$ s$^{-1}$  below $\sim$10 K (close to the charge
order), irrespective of the spin fluctuation rate $\nu_{\rm D}$ 
($\ge10^9$ s$^{-1}$) that is independent of 
temperature as inferred from the previous LF-\msr\ study\cite{Koda:04}. 
The non-linear behavior in the $K$-$\chi$ plot might be thus understood as a dynamical
change of $A_\mu^z$ induced by the slowing down of valence fluctuation.
It is also true, however, that \lvo\ is a metallic system
and thereby the above interpretation would be an oversimplified view of the
actual situation.

The non-linear behavior of $K$-$\chi$ relation is occasionally found in various
magnetic systems including pyrochlore oxides\cite{Dunsiger:03}, where the anomaly 
is attributed to impurities or muon-induced local perturbation of electronic states
(i.e., muons serving as impurities).
However, we presume that such effect 
would be negligibly small in metallic \lvo\ because of the following reasons.
First, the positive charge brought by muon is screened by conduction
electrons in metallic systems so that the local modification of the electronic
state would be negligible. Secondly, it happens that there 
is a large space available in the spinel structure to
accommodate muons; as mentioned earlier, muons are likely sitting
near the center of large octahedron cages cornered by oxygen atoms (see
Figure 5). Then the change in the elastic energy due to accommodation of muon
would be small, although there is no quantitative estimation at this stage.

Finally, we briefly examine the consistency of our \msr\ result with
those obtained by other microscopic probes.
An apparent discrepancy seems to arise from the fact that $^{51}$V/$^7$Li
nuclear magnetic resonance (NMR) reports only a single frequency 
component\cite{Trinkl:00,Kaps:01,Onoda:97,Mahajan:98,Fujiwara:98,Fujiwara:99}, 
whereas at least two components were observed in \msr.  As discussed in the previous 
report\cite{Koda:04}, however, this can be readily understood by the difference
in the frequency window of sensitivity.  It happens that the relaxation rate of the
satellite peak ($\Lambda_1$) is of the order of $\sim1$ MHz, which might mask the
NMR signal. 
The appearance of $\Delta A_\mu^z$ and associated non-linear behavior of
$K$-$\chi$ in the main peak 
might be irrelevant for NMR as long as $\nu_{\rm C}\gg 10^6$ s$^{-1}$
so that the hyperfine coupling may be averaged over the time scale of NMR.
Concerning the results from neutron scattering, the most recent report suggests
the presence of two components with different dynamical characters\cite{Murani:04}. 
Thus, it seems that the ground state of \lvo\ revealed by \msr\ is
consistent with the current experimental knowledge on this compound.

In summary, we have demonstrated that the ground state of \lvo\ is characterized
by the staggered magnetism at low temperatures, where the presence of local
magnetic moments has been established.  While such a ground state may be
explained by the valence fluctuating vanadium ions close to localization,
it is inconsistent with the prediction of conventional heavy-fermion state
based on the Kondo scenario. In order to reconcile the metallic behavior 
as a bulk property with that revealed by the present \msr\ study, further
progress in the theoretical understanding is clearly needed.

We acknowledge comments on our first draft from D. C. Johnston, and 
would like to thank the staff of TRIUMF for technical support during
\msr\ experiment. This work was partially supported by a Grant-in-Aid
for Creative Scientific Research from the Ministry of Education, Culture,
Sports, Science and Technology, Japan. One of the authors (A.K.) was
supported by JSPS.

\newpage

\end{document}